# Multi-quantum-channel mediated tunable single-photon skyrmions from metasurfaces

Yan Wang[1,†], Zhenyu Guo[2,†], Minggui Liang[1], Shuangchun Wen[1],
Yijie Shen[2,3,*], and Hailu Luo[1,*]

[1]*Laboratory for Spin Photonics, School of Physics and Electronics, Hunan University, Changsha 410082, China*

[2]*Centre for Disruptive Photonic Technologies, School of Physical and Mathematical Sciences, Nanyang Technological University, Singapore 637371, Singapore*

[3]*School of Electrical and Electronic Engineering, Nanyang Technological University, Singapore 639798, Singapore*

[†]These authors contributed equally: Yan Wang, Zhenyu Guo

[*]Corresponding email: yijie.shen@ntu.edu.sg; hailuluo@hnu.edu.cn

**Abstract**

Quantum optical skyrmions, as topologically robust quantum information carriers, hold transformative potential for resilient high-dimensional quantum information networks. However, their practical exploitation was still restricted to a single quantum channel, which precludes the multiplexing essential for practical high-capacity quantum networks. Here, we utilize a metasurface to achieve multi-channel quantum state distribution of the polarization-entangled photon pairs, inducing a two-photon bunching effect in both the spin and spatial dimensions with compact flat optics. At the spatial bunching port, controlled manipulation of the spin-orbit interaction enables the generation of a tunable single-photon skyrmion pair. In contrast to any prior skyrmion generation, the single-photon skyrmions are mediated and topologically controlled by quantum measurement in multiple channels. Concurrently, during the amplitude and phase modulation process, both the skyrmion localization and the texture helicity can be precisely customized. The proposed tunable single-photon skyromions offer multidimensional controllability and topological stability provide a viable path toward noise-resilient high-dimensional quantum information processing.

## Introduction

Optical skyrmions[1-6] harness the intrinsic vector nature of light fields to weave complex topological textures, offering an exciting avenue for high-density optical information processing[7-14]. When such vector topology is encoded in single-photon[15-18] or two-photon[19-21] degrees of freedom, optical skyrmions can be extended into the quantum regime. In semiconductor cavity quantum electrodynamics systems, single-photon skyrmion states are stably generated by engineering the spin-orbit coupling of quantum emitters via micro/nano structures[22-25], thereby providing integrated high-dimensional quantum light sources for advanced quantum photonic information technologies. On the other hand, studies have confirmed that the topological invariant of quantum skyrmions exhibits unprecedented robustness during transport, positioning them as promising ideal carriers for quantum information processing in complex environments[26-30]. However, current research on quantum skyrmions is generally restricted to a limited number of quantum channels, leading to significant technical limitations in core application scenarios such as high-capacity multiplexed transmission of multi-degree-of-freedom quantum information and high-bit topological optical quantum computing. Consequently, realizing multi-channel generation and multi-dimensional control of quantum skyrmions within a compact space remains a formidable challenge.

Metasurfaces[31-34], as two-dimensional artificial structures composed of arrays of subwavelength-scale nanoantennas, possess the ability to precisely control the amplitude, phase, and polarization of optical fields with multiple degrees of freedom (DOFs)[35-44], thus providing an effective means to address this challenge. The geometric phase gradient on a Pancharatnam-Berry (PB) phase metasurface imparts a spin-dependent transverse momentum to the left- and right-handed circularly polarized (LCP and RCP) components, causing their propagation paths to deviate from the initial direction[45-47]. Building on this foundation, the incorporation of a dynamic phase for joint manipulation of the optical field enables spin decoupling with a single optical element, thereby establishing a new framework for the flexible reconstruction and steering of optical skyrmions[48]. Moreover, by constructing meta-fibers from polarization-dependent metasurfaces supported on the tip of an optical fiber, diverse optical skyrmions can be generated on demand with simplicity and efficiency within a compact space[49]. The outstanding capability of metasurfaces for generating and controlling skyrmions points to the possibility of devising a novel type of quantum device; nevertheless, the detailed realization scheme remains to be explored.

In this article, we generate a pair of tunable single-photon skyrmions based on a PB phase metasurface, whose topological textures and skyrmion numbers can be precisely controlled. The core concept behind the realization of tunable single-photon skyrmions is to achieve multi-channel distribution of entangled states via a metasurface, and to tailor symmetric topological structures in the two entangled photons, with skyrmion numbers of +2 and −2, respectively. The inherent multi-channel nature of the metasurface enables functional exchange and role reversal between the signal and idler photons, providing a unique means for manipulating tunable single-photon skyrmions. In the two diffraction-order channels of opposite polarizations in the metasurface, the skyrmion number is controlled by switching the projection basis. On this basis, by tuning the amplitude ratio and phase difference between the orthogonal spin components, the skyrmion size and texture helicity can be precisely engineered. Unlike conventional single-photon skyrmions, the skyrmion number of this tunable single-photon skyrmion pair can be controlled by orthogonal polarization projections of the entangled state, which greatly enhances their potential as reconfigurable quantum information carriers. Our work combines the unique advantages of polarization entanglement and metasurfaces, offering a promising avenue for topological engineering and modulation in quantum information processing.

## Results

### Theoretical framework

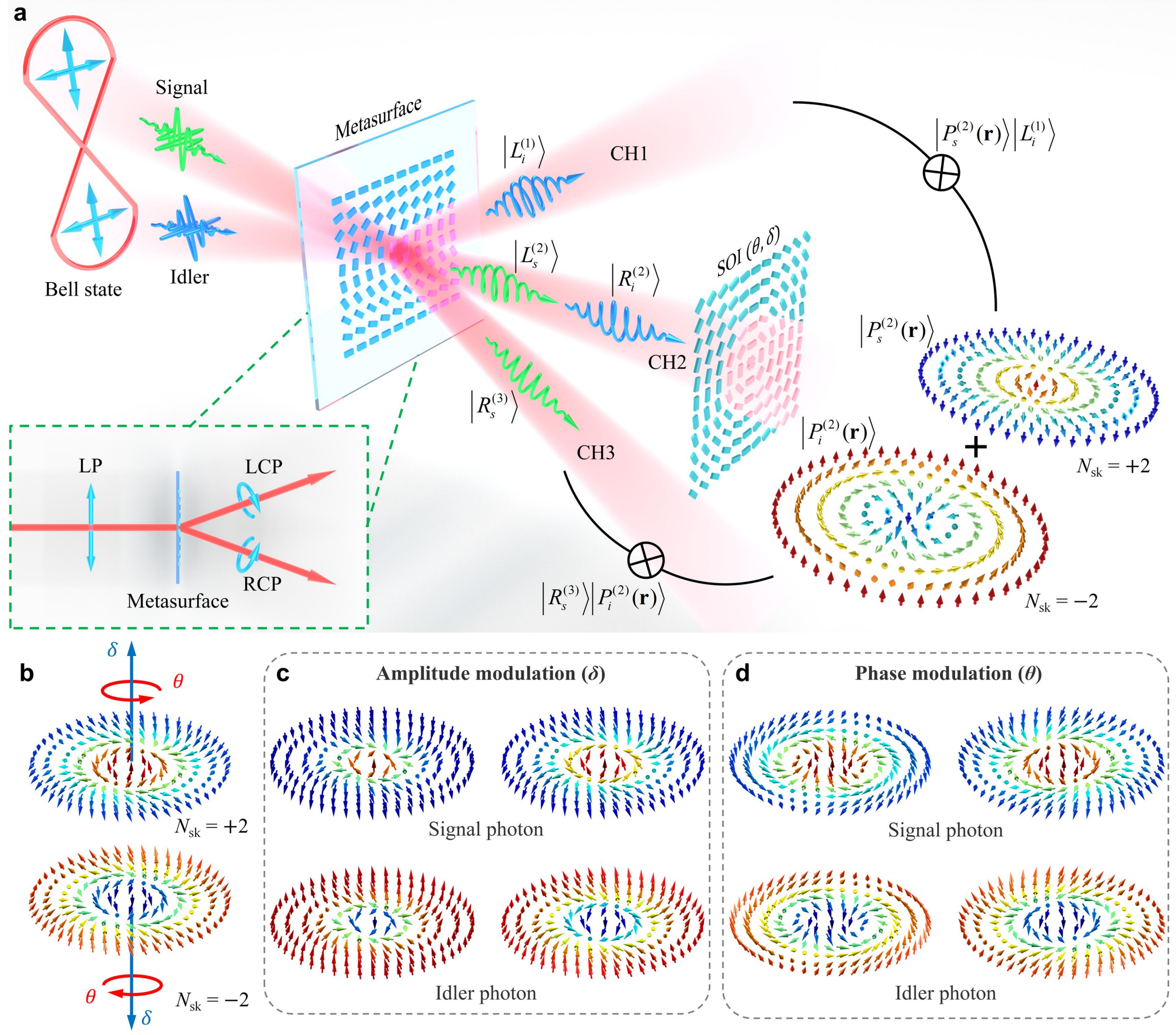

**Figure 1. Schematic illustration of the multi-channel generation and manipulation of the tunable single-photon skyrmions. a,** A PB phase metasurface enables the formation of tunable single-photon skyrmions. Under coincidence measurement, the LCP (RCP) component of the signal (idler) photon undergoes a tunable SOI simultaneously within channel CH2 of the metasurface, generating an optical skyrmion with a positive (negative) skyrmion number. Inset: schematic illustration of the transmission of linearly polarized light through a metasurface. **b,** Independent modulation of the amplitude and phase of tunable single-photon skyrmions via SOI. **c** and **d,** Manipulation of the vector field through the two degrees of freedom of amplitude and phase.

As shown in the inset of Fig. 1a, the metasurface that generates a PB phase gradient constitutes the core element for constructing the tunable single-photon skyrmions. It imparts a half-wave birefringent phase retardation, while its spatially anisotropic optical axis distribution is described by $\eta(x,y) = \pi x / \Lambda$, where $\Lambda$ denotes the spatial period of the local optical-axis variation. Consequently, the photonic spin Hall effect (SHE) on metasurface induces a spin-dependent evolution of the single-photon polarization wave function[46]:

$$|\psi\rangle \to \hat{U}|\psi\rangle = \exp[-i\Phi_{\text{PB}}(x,y)]\hat{P}_{LR}|\psi\rangle + \exp[i\Phi_{\text{PB}}(x,y)]\hat{P}_{RL}|\psi\rangle. \tag{1}$$

Here, $\hat{U}$ denotes the transformation operator of the metasurface acting on an arbitrary quantum state $|\psi\rangle$, while $\hat{P}_{LR} = |R\rangle\langle L|$ and $\hat{P}_{RL} = |L\rangle\langle R|$ are the projection operators associated with the two diffraction orders of the metasurface. The geometric phase acquired by the photon is given by $\Phi_{\text{PB}}(x,y) = -2\eta(x,y) = -2\pi x/\Lambda$. Owing to the additional phase gradient $\nabla\Phi_{\text{PB}}$ introduced along the $x$ direction, the photon acquires a spin-dependent transverse momentum shift $\Delta\mathbf{k} = -\sigma_{\pm}\nabla\Phi_{\text{PB}} = \mp(2\pi/\Lambda)\hat{\mathbf{e}}_x$ after passing through the metasurface. Therefore, in the far-field region of diffraction, the LCP and RCP components are completely spatially separated, allowing the output polarization state to be directly determined through position measurements.

Within a small range of incident angles, the half-wave birefringent phase retardation required by Eq. (1) can be approximately satisfied, which permits oblique incidence of the photons onto the metasurface at an incident angle equal to the $\pm1$ st-order diffraction angle[50,51]. As illustrated in Fig. 1a, a pair of polarization-entangled photons is obliquely incident onto the metasurface to realize the two-photon SHE[47]. In our configuration, the entangled photon pair is prepared in a superposition of the third and fourth Bell states, $|\Psi^{+}\rangle$ and $|\Psi^{-}\rangle$ [52]:

$$|\Psi_{\text{in}}\rangle = \frac{1}{2}\left(1+e^{i\gamma}\right)|\Psi^{+}\rangle + \frac{1}{2}\left(1-e^{i\gamma}\right)|\Psi^{-}\rangle = \frac{1}{\sqrt{2}}\left(|H_s\rangle|V_i\rangle + e^{i\gamma}|V_s\rangle|H_i\rangle\right), \tag{2}$$

where the subscripts $s$ and $i$ denote the signal and idler photons, respectively, and $\gamma$ represents the relative phase between the two photons. After interacting with the metasurface, the output state of the entangled photon pair is calculated to be:

$$\begin{aligned}|\Psi_{\text{out}}\rangle = \hat{U}_s\hat{U}_i|\Psi_{\text{in}}\rangle = -\frac{i\left(1+e^{i\gamma}\right)}{2\sqrt{2}}\left[e^{-4i\eta(x,y)}|R_s^{(3)}\rangle|R_i^{(2)}\rangle - e^{4i\eta(x,y)}|L_s^{(2)}\rangle|L_i^{(1)}\rangle\right] \\ -\frac{i\left(1-e^{i\gamma}\right)}{2\sqrt{2}}\left[|L_s^{(2)}\rangle|R_i^{(2)}\rangle - |R_s^{(3)}\rangle|L_i^{(1)}\rangle\right].\end{aligned} \tag{3}$$

Here, $\hat{U}_s$ and $\hat{U}_i$ denote the transformation operators of the metasurface acting on the signal and idler photons, respectively, and the superscripts (1), (2), and (3) indicate that the photons reside in channels CH1, CH2, and CH3, respectively. Equation (3) demonstrates the spin bunching and anti-bunching phenomena of the two photons on the metasurface: when the incident photon pair is in a maximally entangled state ($\gamma = 0$ and $\pi$), the signal and idler photons always collapse into identical or opposite spin states, and the correlation in the bunching behavior can be arbitrarily tuned via the relative phase $\gamma$. Under the condition of strict wavelength degeneracy between the signal and idler photons, the spin bunching effect enables the realization of an asymmetric linear optical network with two input ports and three output ports on the metasurface. Among the three output channels, the photons in CH1 and CH3 can be unambiguously identified as $|L_i\rangle$ and $|R_s\rangle$, respectively. However, in CH2, the $|L_s\rangle$ and $|R_i\rangle$ components are transmitted in the same direction, so that the polarization state of the photons in CH2 cannot be determined by single-channel intensity measurements alone. It is precisely this spatial indistinguishability that allows the selective capture of photons with a specified polarization state from CH2 by performing coincidence measurements. For instance, when coincidence measurements are carried out between CH1 and CH2, only the $|L_s\rangle$ component in CH2 contributes to the coincidence counts, which directly quantifies the probability of spin bunching of the entangled photon pair occurring on the metasurface. It is worth noting that the roles of the signal and idler photons are no longer predetermined, i.e., the photon used for heralding and spatial quantum information processing can be interchanged.

To generate tunable single-photon skyrmions based on the metasurface, additional optical elements were inserted into CH2 to induce specific SOI, thereby converting circularly polarized Gaussian-mode photons into Stokes vector skyrmions with topological invariants of +2 and −2, respectively. In CH2, the initial polarization states of the signal and idler photons are unambiguously identified as LCP and RCP states, respectively; upon undergoing the SOI, these photons are transformed into (see Supplementary Note 1 for more details):

$$
\begin{aligned}
\left|L_s^{(2)}\right\rangle &\rightarrow \left|P_s^{(2)}(r,\varphi)\right\rangle = \cos\frac{\delta}{2}\mathrm{LG}_{+20}(r,\varphi)\left|R_s^{(2)}\right\rangle - i\sin\frac{\delta}{2}e^{i2\theta}\mathrm{LG}_{00}(r,\varphi)\left|L_s^{(2)}\right\rangle, \\
\left|R_i^{(2)}\right\rangle &\rightarrow \left|P_i^{(2)}(r,\varphi)\right\rangle = \cos\frac{\delta}{2}\mathrm{LG}_{-20}(r,\varphi)\left|L_i^{(2)}\right\rangle - i\sin\frac{\delta}{2}e^{-i2\theta}\mathrm{LG}_{00}(r,\varphi)\left|R_i^{(2)}\right\rangle.
\end{aligned} \tag{4}
$$

Here, $\mathrm{LG}_{lp}(r,\varphi)$ denotes the electric field expression of a Laguerre-Gaussian (LG) beam with a vortex topological charge of $l$ and a radial index of $p$, where the spatial coordinates $r$ and $\varphi$ represent the radial distance and polar angle in the plane polar coordinate system, respectively. Equation (4) demonstrates that the signal and idler photons give rise to a pair of vector vortex fields with opposite topological charges, denoted as $\left|P_s^{(2)}(r,\varphi)\right\rangle$ and $\left|P_i^{(2)}(r,\varphi)\right\rangle$ within CH2. Their topological textures can be characterized by measuring the Stokes parameters (see Methods for more details). The two-photon wave function with skyrmion mode can be obtained by substituting Eq. (4) into Eq. (3):

$$
\begin{aligned}
\left|\Psi_{\mathrm{sk}}(r,\varphi)\right\rangle = &-\frac{i\left(1+e^{i\gamma}\right)}{2\sqrt{2}}\left[e^{-4i\eta(x,y)}\left|R_s^{(3)}\right\rangle\left|P_i^{(2)}(r,\varphi)\right\rangle - e^{4i\eta(x,y)}\left|P_s^{(2)}(r,\varphi)\right\rangle\left|L_i^{(1)}\right\rangle\right] \\
&-\frac{i\left(1-e^{i\gamma}\right)}{2\sqrt{2}}\left[\left|P_s^{(2)}(r,\varphi)\right\rangle\left|P_i^{(2)}(r,\varphi)\right\rangle - \left|R_s^{(3)}\right\rangle\left|L_i^{(1)}\right\rangle\right].
\end{aligned} \tag{5}
$$

Therefore, coincidence measurements performed between CH1 and CH2 and between CH2 and CH3 yield the polarization projections $\left|R_s^{(3)}\right\rangle\left|P_i^{(2)}(r,\varphi)\right\rangle$ and $\left|P_s^{(2)}(r,\varphi)\right\rangle\left|L_i^{(1)}\right\rangle$, respectively, with both projection probabilities given by $P(\gamma) = \left|-ie^{\pm 4i\eta(x,y)}\left(1+e^{i\gamma}\right)/\left(2\sqrt{2}\right)\right|^2 = (1/4)\left(1+\cos\gamma\right)$. To maximize the projection probability, the relative phase $\gamma$ is set to $0$, corresponding to the incident entangled photon pair being prepared in the third Bell state, for which the projection probability becomes $P(0) = 1/2$. On this basis, by adjusting the action of the SOI implemented in CH2, the amplitude parameter $\delta$ and the phase factor $\theta$ in Eq. (4) can be precisely controlled, thereby enabling independent modulation of the skyrmion size and texture helicity. Under the initial conditions of $\delta = \pi/2$ and $\theta = 0$, the signal and idler photons respectively generate the tunable single-photon skyrmions as illustrated in Fig. 1b. Owing to the perfectly collinear propagation of the signal and idler photons, the skyrmions created by the two opposite spin states completely overlap in space, such that at every spatial point the two skyrmions occupy antipodal points on the Poincaré sphere. In the amplitude modulation process shown in Fig. 1c, the skyrmion size is altered, whereas the texture helicity is preserved. As shown in Fig. 1d, during phase modulation, the vector field undergoes a global in-plane rotation while the skyrmion size remains unchanged.

The metasurface mixes the transmission paths of the signal and idler photons, enabling their functional roles to be interchanged. This breaks the inherent limitation in polarization-entanglement-based nonlocal quantum information processing, where the identities of the photons used for spatial information processing and heralding are fixed. For the signal photon, the LCP and RCP components are respectively employed for spatial information processing with a positive topological charge and as a heralding signal for the idler photon. The situation with idler photons is exactly the opposite of that with signal photons. Consequently, the tunable single-photon skyrmions coexist within the spatial information output channel of the metasurface, and the polarized projection associated with the signal photon or the idler photon can be selectively observed by switching the heralding channel.

**Experimental setup**

As shown in Fig. 2a, polarization-entangled photon pairs are generated via a type-II spontaneous parametric down-conversion (SPDC) process[53] in a 20-mm-long periodically poled $KTiOPO_4$ (PPKTP) crystal with grating period of 10.025 $\mu$m (see Methods for more details). Since the half-wave phase retardation of the metasurface is actually designed for normal incidence, a narrow-linewidth pump source is employed to reduce the spectral linewidth of the down-converted photons, thereby mitigating the additional dispersion and waveform distortion induced by the oblique incidence of light onto the metasurface. Here, the central wavelength and spectral linewidth of the pump source are 405.018 nm and 1.5 MHz, respectively. To ensure wavelength degeneracy, the PPKTP crystal is operated at a temperature of 23.787°C, which provides an optimal match between its grating period and the central wavelength of the pump laser. As shown in Fig. 2b, polarization correlation curves are measured in

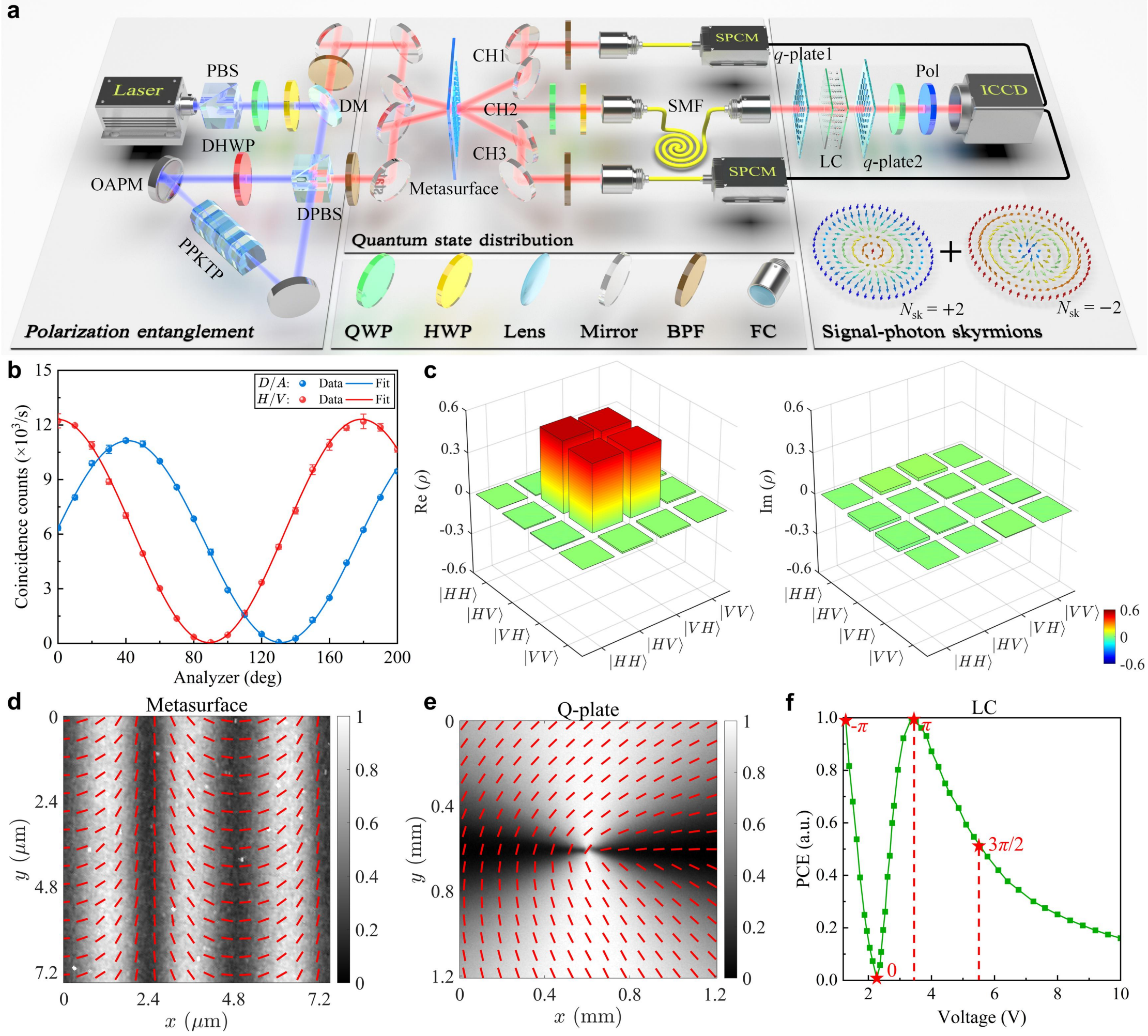

**Figure 2. Experimental setup for the multi-channel generation and manipulation of tunable single-photon skyrmions using a metasurface. a,** Experimental configuration. Laser, 405-nm narrow-linewidth continuous-wave laser; QWP, quarter-wave plate; HWP, half-wave plate; DM, dichroic mirror; DPBS, dual-wavelength polarization beam splitter; DHWP, dual-wavelength half-wave plate; PPKTP, periodically pole $KTiOPO_4$ crystal; OAPM, off-axis parabolic mirror; BPF, band-pass filter; FC, fiber coupler; SMF, single-mode fiber; $q$-plate, half-wave phase-retardation spiral phase plate with $q = 0.5$; SPCM, single-photon counting module; LC, liquid-crystal phase shifter; Pol, linear polarizer; ICCD, intensified charge-coupled device. **b,** Polarization correlation curves measured in the $D/A$ and $H/V$ measurement bases. **c,** Density matrix reconstructed via quantum state tomography. **d** and **e,** Optical-axis distributions of the metasurface and the $q$-plate reconstructed from normalized intensity images under cross-polarized microscopy. **f** Modulation curves of the amplitude ratio for circularly polarized light imparted by the LC phase shifter.

the $D/A$ and $H/V$ bases, yielding interference visibilities of $\mathcal{V}_{D/A} = (99.31 \pm 0.24)\%$ and $\mathcal{V}_{H/V} = (99.11 \pm 0.35)\%$, respectively, which significantly exceed the threshold of $1/\sqrt{2} \approx 70.7\%$ for violating Bell's inequality[54,55] (see

Supplementary Note 2 for more details). The near-unity visibilities attest to the exceptionally high degree of entanglement of the photon pair in both the phase and amplitude DOFs. Subsequently, a quantum state tomography[56] (QST) measurement is performed to reconstruct the density matrix, which exhibited a fidelity of $F = 0.988 \pm 0.011$ (Fig. 2c), further confirming the overall excellent entanglement quality of the prepared polarization-entangled state.

In this work, the PB phase metasurface and the $q$-plate are the key elements for generating tunable single-photon skyrmions. These two optical components are fabricated by writing spatially varying nanogrooves into fused silica using a femtosecond laser (see Methods for more details). In experimental characterization, scanning electron microscopy (SEM) is commonly employed to accurately characterize the metasurfaces and $q$-plates, whereby the optical-axis distribution can be fully determined by observing the nanogrooves. However, to avoid permanent damage during characterization process, the optical-axis distribution can alternatively be reconstructed from the normalized intensity distribution $I(x, y)$ acquired under a cross-polarized microscope (see Supplementary Note 3 for more details). As shown in Figs. 2d and 2e, the optical-axis distributions of the metasurface and the $q$-plate are in excellent agreement with the design parameters.

To enable accurate manipulation of the tunable single-photon skyrmions, the amplitude ratio and the phase difference between LCP and RCP photons can be modulated via an LC phase shifter. Here, the LC phase shifter can generate a birefringent phase delay that is precisely controlled by an external voltage. According to the transformation relation between the linear and circular polarization bases, a phase difference between two orthogonal linear polarization states is equivalent to an amplitude ratio between the two circular polarization states. To verify the amplitude-modulation capability of the LC phase shifter for circularly polarized light, we placed it between two crossed circular polarizers and recorded the polarization conversion efficiency (PCE) as a function of the phase shift, as shown in Fig. 2f (see Supplementary Note 4 for more details). Here, the phase retardation imparted by the liquid crystal is driven by an external voltage. In Fig. 2f, the four special points marked by red stars possess phase retardations of $-\pi$, 0, $\pi$, and $3\pi/2$, respectively.

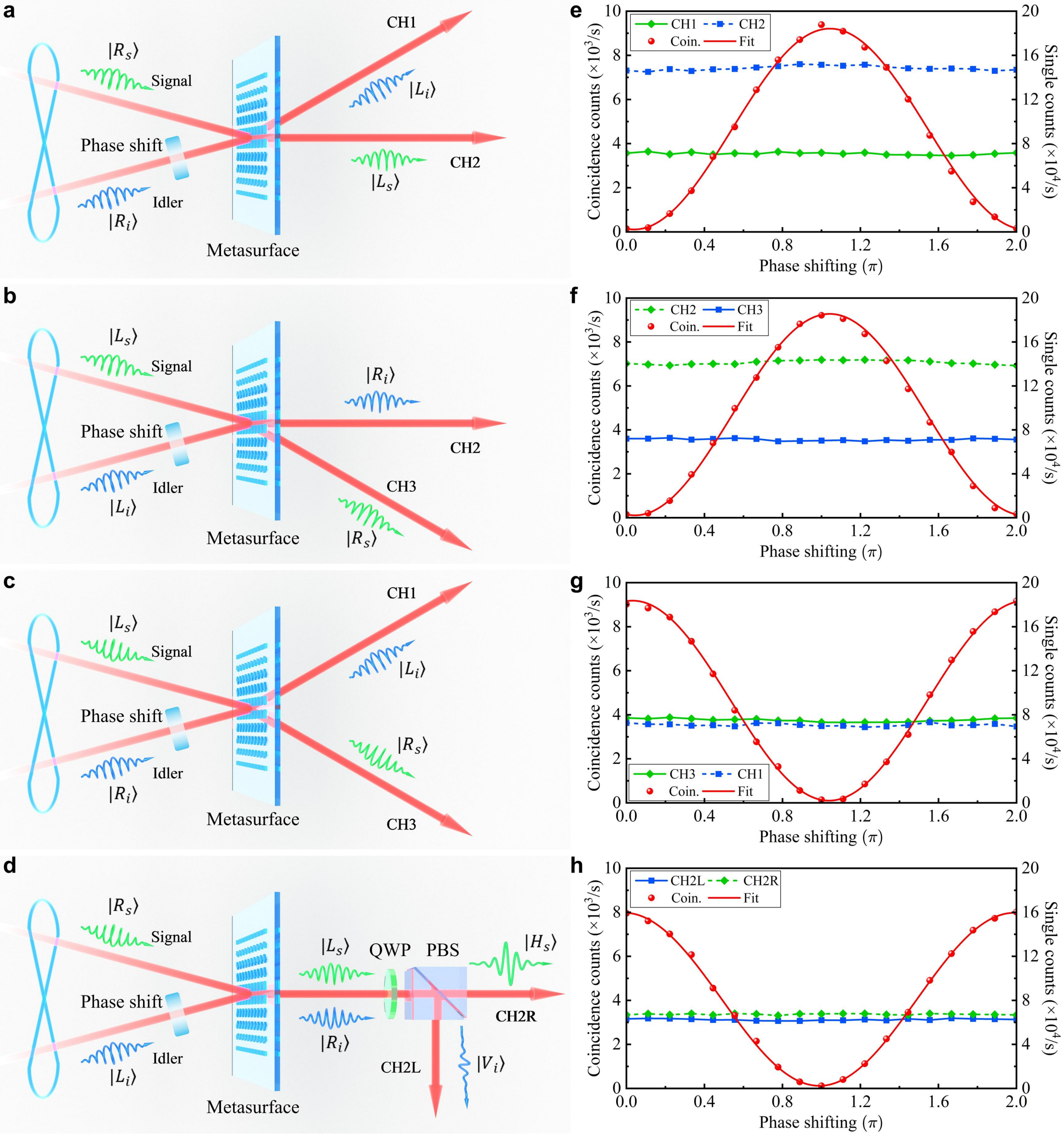

**Figure 3. Experimental results of spin bunching and anti-bunching on the metasurface. a-d,** Transmission scenarios for different spin projection combinations of the signal photon (green) and the idler photon (blue) on the metasurface. **e-h,** Experimental results of single-channel counts and coincidence measurements for the configurations shown in (a)-(d), respectively. The green and blue curves represent the single-channel counts of the signal and idler photons over one second, with solid and dashed lines denoting the incident photons in the LCP and RCP states, respectively; the red curve represents the coincidence counts per second.

### Experimental measurement of quantum state distribution on the metasurface

To demonstrate the distribution of multi-channel quantum states on the metasurface,

the polarization correlation curves are characterized via coincidence measurements at the output ports. As presented in Figs. 3a-3d, the signal photon (shown in green) and the idler photon (shown in blue) in a polarization-entangled state are respectively injected onto the metasurface at the $\pm 1$ st-order diffraction angles, and single-channel counts as well as coincidence counts within one second are recorded at the three output ports. Since both photons possess two orthogonal spin components, there are a total of four possible output states. Taking Fig. 3a as an example, both signal photons and idler photons undergo a spin state reversal and acquire additional transverse momentum. In this scenario, it can be unambiguously determined that both the output signal and idler photons are converted into the LCP state and emerge from CH2 and CH1, respectively. However, in the situation depicted in Fig. 3d, the paths of the signal and idler photons are completely overlapped. Consequently, a combination of a QWP at $-45°$ and a PBS is required to further separate their propagation paths, thereby collecting the signal photon and the idler photon in channels CH2R and CH2L, respectively.

Subsequently, a birefringent phase shifter is introduced into the idler arm before the metasurface to actively control the polarization correlation between the two photons. Throughout the phase-shifting process, the single-channel counts remain approximately constant as a horizontal line, whereas the coincidence measurements demonstrate that the probabilities of spin bunching (Figs. 3e and 3f) and anti-bunching (Figs. 3g and 3h) exhibit sinusoidal variations with opposite phases. By performing sinusoidal fitting to the experimental data points, polarization interference curves are obtained for the four cases. The corresponding polarization interference visibilities are calculated as $\mathcal{V}_{12} = (98.10 \pm 0.46)\%$ , $\mathcal{V}_{23} = (98.92 \pm 1.16)\%$ , $\mathcal{V}_{31} = (98.18 \pm 1.35)\%$, and $\mathcal{V}_{22} = (97.54 \pm 0.51)\%$, respectively. In our configuration, the visibilities of these four interference curves are mainly determined by the entanglement quality of the photon pair in the phase DOF and the polarization conversion efficiency of the metasurface. Owing to the inevitable dispersion and waveform distortion induced by the oblique incidence of photons onto the metasurface, the recorded coincidence counts show relatively large fluctuations, leading to appreciable uncertainties in the measured polarization interference visibilities. Technically, this can be mitigated by further narrowing the pump laser linewidth, reducing the length of the PPKTP crystal, or incorporating an additional Fabry-Pérot cavity in free space before the metasurface for spatial filtering. The experiment demonstrates that, by modulating the relative phase between the incident entangled photon pairs, continuous control of two-photon spin and spatial bunching on the metasurface can be achieved without altering the single-channel counts.

**Generation and modulation of tunable single-photon skyrmions**

Based on the analysis of Eq. (5), the tunable single-photon skyrmions with the strongest correlations can be obtained by inputting an entangled photon pair in the third Bell state onto the metasurface. As shown in Fig. 2a, in the output channels CH3 and CH1 of the metasurface, the RCP component of the signal photon and the LCP component of the idler photon are respectively collected via fiber couplers and single-mode fibers (SMFs), and the single-photon counts per second are recorded by single-photon counting modules (SPCMs). In channel CH2, the LCP component of the signal photon and the RCP component of the idler photon propagate collinearly and are simultaneously coupled into an 18-m-long SMF, to compensate for the extra time delays arising from the photoelectric conversion within the SPCMs and the transmission of the electrical signals to the control port of the intensified charge-coupled device (ICCD). Here, the ICCD is set to external trigger mode for coincidence imaging, meaning the shutter opens only for a brief window of time when a trigger signal is received (see Methods for more details). After the photons in CH2 are output again as fundamental Gaussian spatial modes, the LCP signal photon and the RCP idler photon undergo opposite SOI on the $q$-plate1, thereby converting their spatial modes into two LG modes with opposite topological charges. Subsequently, by controlling the external voltage and optical axis orientation of the LC phase shifter, the signal photon and the idler photon are converted into two orthogonal elliptical polarization states. After the spin-dependent evolution on the $q$-plate2, the LCP and RCP components of each individual photon coherently superpose, thereby generating the tunable single-photon skyrmions at the output port.

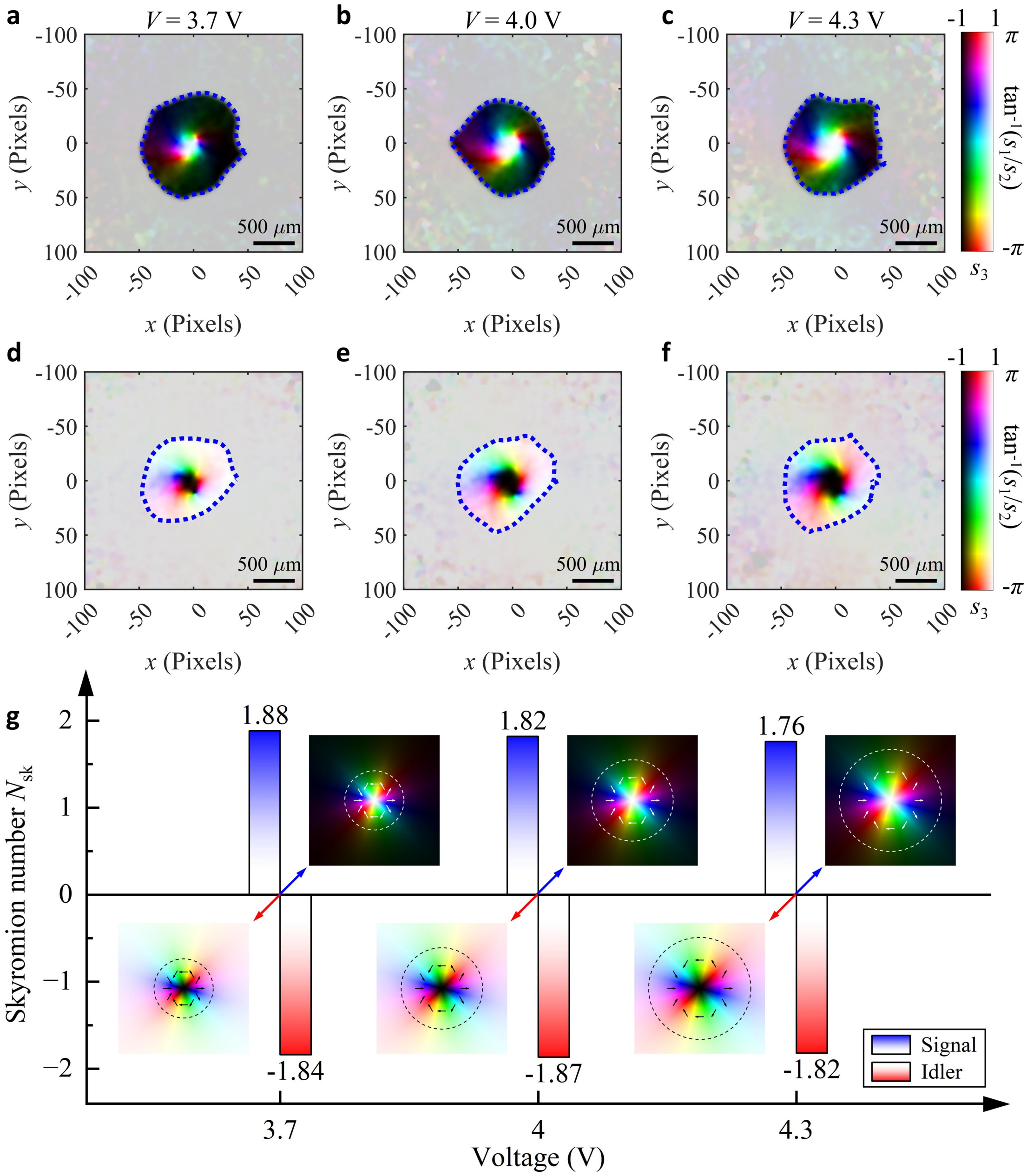

**Figure 4. Amplitude modulation results of the tunable single-photon skyrmions.** **a-c,** Experimental results on the amplitude modulation of skyrmions for the signal photon. **d-f,** Experimental results on the amplitude modulation of skyrmions for the idler photon. **g,** The skyrmion numbers calculated using the snake algorithm.The inset shows the theoretical variation of skyrmion size with amplitude modulation.

To experimentally characterize the vector field of the tunable single-photon skyrmion, the Stokes parameters are retrieved by jointly using a QWP and a linear polarizer (Pol) (see Supplementary Note 5 for more details). The two-photon coincidence measurements are first performed with the angle fixed at $\theta = 0$, and the vector field of the signal photon is observed by selecting CH1 as the heralding channel. As shown in Figs. 4a-4c, during the amplitude modulation, the central intensity of the vector field gradually increases, signifying that the Gaussian mode progressively becomes dominant. To calculate the skyrmion number according to Eq. (8), the optimal integration region identified using the snake algorithm[23,29] is indicated by the blue dashed lines in Fig.4a-4c. As shown by the blue bars in Fig. 4g, the skyrmion numbers of signal photon are calculated to be $N_{\mathrm{sk}}(V = 3.7\ \mathrm{V}) = +1.88$ , $N_{\mathrm{sk}}(V = 4.0\ \mathrm{V}) = +1.82$ , and $N_{\mathrm{sk}}(V = 4.3\ \mathrm{V}) = +1.76$ . The slight deficit of these values compared with the theoretical skyrmion number $N_{\mathrm{sk}} = +2$ is attributed to the finite degree of polarization entanglement, fabrication imperfections of the metasurface, and the center misalignment between the two $q$-plates.

The skyrmion generated by the idler photon and that generated by the signal photon are topologically correlated, a relationship that originates from the projection of the two-photon entangled state onto different polarization bases. To observe the vector field of the idler photon, the heralding channel is switched from CH1 to CH3, thereby producing a skyrmion whose polarization distribution is everywhere orthogonal to that of the signal photon (Figs. 4d-4f). As indicated by the red bars in Fig. 4g, the skyrmion numbers of the idler photon are computed to be $N_{\mathrm{sk}}(V = 3.7\ \mathrm{V}) = -1.84$, $N_{\mathrm{sk}}(V = 4.0\ \mathrm{V}) = -1.87$ , and $N_{\mathrm{sk}}(V = 4.3\ \mathrm{V}) = -1.82$ , which are close to the theoretical value of $N_{\mathrm{sk}} = -2$. These values approximately satisfy an opposite-sign relationship with the skyrmion numbers of the signal photon, which is consistent with the theoretical expectation. The results presented in Fig. 4 demonstrate that amplitude modulation of the tunable single-photon skyrmion can be realized via the LC phase shifter, while the skyrmion number remains topologically robust throughout the modulation.

By rotating the optical axis orientation of the LC while keeping the external voltage constant, phase modulation between the circularly polarized components is realized, resulting in a global rotation of the skyrmion vector field. As shown in Figs. 5a-5c, the overall intensity of the signal-photon skyrmion remains constant, whereas its phase distribution undergoes a rotation that is positively correlated with the rotation angle $\theta$. For the idler photon, the phase distribution rotates in the same direction, and at every spatial point it maintains exactly the same phase value as that of the signal photon (Figs. 5d-5f). As presented in Fig. 5g, throughout the entire phase modulation process, the skyrmion numbers of both the signal and idler photons exhibit excellent robustness, staying close to the theoretical values of +2 and −2, respectively. Consequently, phase modulation of the skyrmion can be realized in a straightforward manner simply by rotating the optical axis of the birefringent device placed between the two $q$-plates.

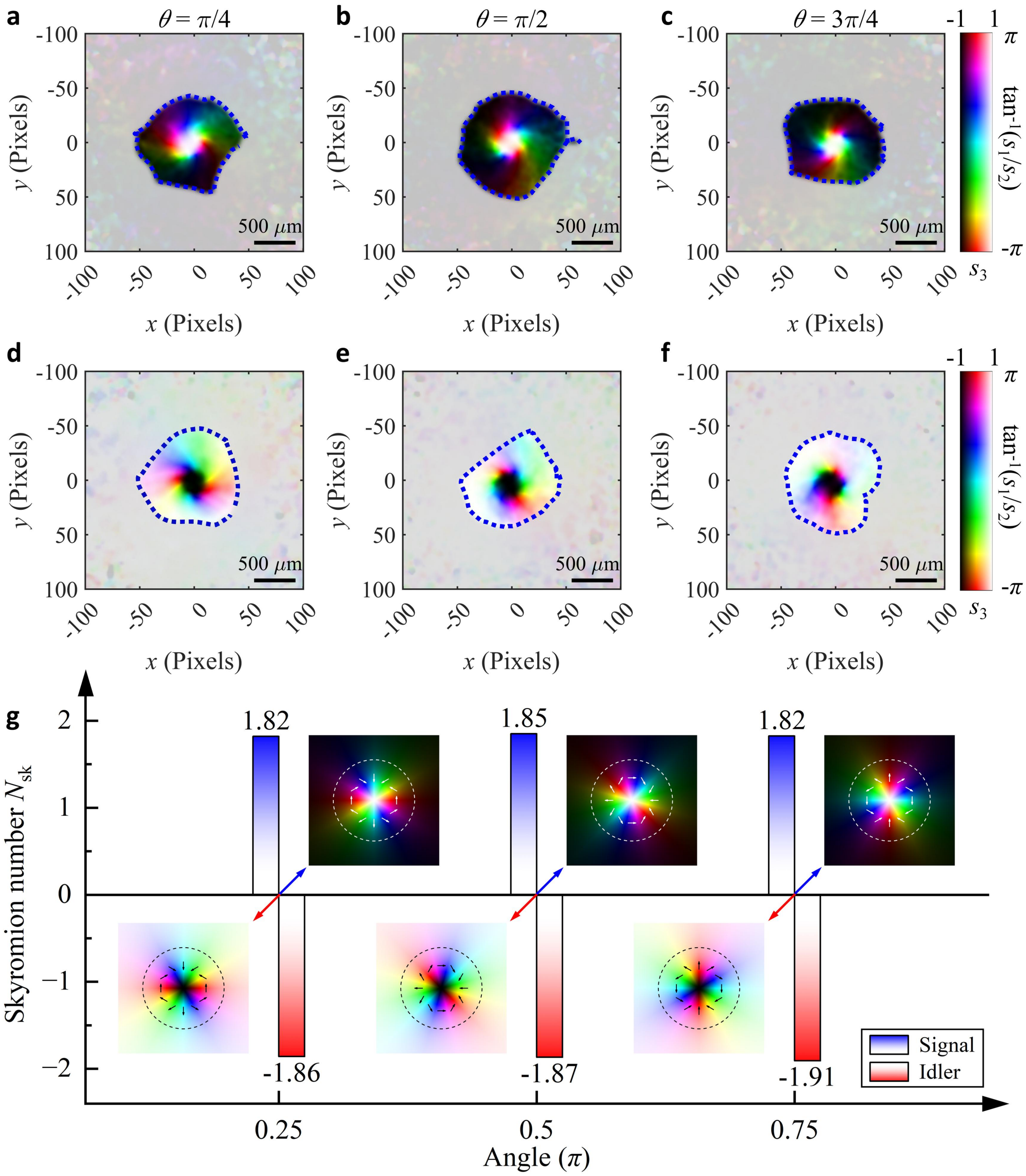

**Figure 5. Phase modulation results of the tunable single-photon skyrmions. a-c,** Experimental results on the phase modulation of skyrmions for the signal photon. **d-f,** Experimental results on the phase modulation of skyrmions for the idler photon. **g,** The inset shows the theoretical variation of texture helicity with phase modulation.

## Discussion

In summary, we exploit the multi-channel capability of a metasurface to generate a pair of topologically tunable single-photon skyrmions from the opposite spin projections of a two-photon entangled state. Unlike previous studies, this skyrmion pair exhibits diverse topological structures in coincidence measurements across different quantum channels, laying the foundation for multi-channel topological quantum state distribution. By injecting polarization-entangled photon pairs into a PB phase metasurface with opposite diffraction orders, we construct a 2×3 linear optical network for multi-channel distribution and tailor tunable topological structures at the metasurface output ports, such that the signal and idler photons respectively form the single-photon skyrmion pairs with symmetric polarization structures and opposite skyrmion numbers. In the amplitude and phase degrees of freedom, the size and texture helicity of the skyrmion pair are symmetrically controlled, allowing them to serve as independent or joint encoding degrees of freedom for multi-parameter high-dimensional encoding on a single photon, thereby directly multiplying the quantum information capacity at the physical level.

Crucially, this tunable topological property, combined with the inherent topological protection of skyrmions, endows the quantum states with robust operation in realistic environments under turbulence and noise. This feature allows quantum key distribution to adopt topologically protected high-dimensional encoding schemes, increasing the key rate while effectively suppressing error rates. Moreover, the scalability of the skyrmion number and its topological protection enable quantum teleportation to directly transfer high-dimensional topological quantum states, maintaining higher fidelity in the presence of noise. By fully leveraging multi-channel distribution and topological correlation detection, quantum secret sharing can realize secure multiparty protocols that are resistant to eavesdropping and capable of detecting internal cheating. These results establish the tunable single-photon skyrmion pair as a unique information carrier of engineered topologically nontrivial entangled states, opening a significant avenue for topology-assisted quantum communication and high-dimensional quantum information processing.

# Methods

### Characterizing skyrmions using the Stokes vector

The Stokes parameters of the signal and idler photons are defined as:

$$\begin{cases} S_0 = |A_k|^2 + |B_k|^2 \\ S_1 = |A_k|^2 - |B_k|^2 \\ S_2 = 2\,\mathrm{Re}\left[A_k^* B_k\right] \\ S_3 = 2\,\mathrm{Im}\left[A_k^* B_k\right] \end{cases}, \; k = s, i, \tag{6}$$

where $A_k$ and $B_k$ represent the horizontal and vertical polarization components, respectively, and are calculated as:

$$\begin{aligned} A_s &= \frac{1}{\sqrt{2}}\left[\cos\frac{\delta}{2}\mathrm{LG}_{+20}(r,\varphi) - i\sin\frac{\delta}{2}e^{i2\theta}\mathrm{LG}_{00}(r,\varphi)\right], \\ B_s &= \frac{-i}{\sqrt{2}}\left[\cos\frac{\delta}{2}\mathrm{LG}_{+20}(r,\varphi) + i\sin\frac{\delta}{2}e^{i2\theta}\mathrm{LG}_{00}(r,\varphi)\right], \\ A_i &= \frac{1}{\sqrt{2}}\left[\cos\frac{\delta}{2}\mathrm{LG}_{-20}(r,\varphi) - i\sin\frac{\delta}{2}e^{-i2\theta}\mathrm{LG}_{00}(r,\varphi)\right], \\ B_i &= \frac{i}{\sqrt{2}}\left[\cos\frac{\delta}{2}\mathrm{LG}_{-20}(r,\varphi) + i\sin\frac{\delta}{2}e^{-i2\theta}\mathrm{LG}_{00}(r,\varphi)\right]. \end{aligned} \tag{7}$$

Subsequently, the three-dimensional Stokes vector constructed from the three normalized parameters $s_1 = S_1 / S_0 = \cos\alpha(\varphi)\sin\beta(r)$, $s_2 = S_2 / S_0 = \sin\alpha(\varphi)\sin\beta(r)$, and $s_3 = S_3 / S_0 = \cos\beta(r)$ as $\mathbf{s} = (s_1, s_2, s_3)$ is employed to characterize the topological texture of the optical field. Here, the angles $\alpha$ and $\beta$ denote the azimuthal and polar angles, respectively, in the spherical coordinate system of the Stokes parameter space; they govern the in-plane and out-of-plane components of the Stokes vector $\mathbf{s}$. This gives rise to a pair of skyrmions with opposite vectorial distributions generated by the signal and idler photons, and the skyrmion number is

calculated as[4]:

$$N_{\mathrm{sk}} = \frac{1}{4\pi}\iint_{\Sigma} \mathbf{s} \cdot \left( \frac{\partial \mathbf{s}}{\partial x} \times \frac{\partial \mathbf{s}}{\partial y} \right) \mathrm{d}x\mathrm{d}y. \tag{8}$$

**Preparation of the polarization-entangled photon pair**

Under the condition of type-II quasi-collinear phase matching, a horizontally polarized high-frequency photon splits into a pair of wavelength-degenerate, orthogonally polarized down-converted photons in a PPKTP crystal, and this process can be described as:

$$\left|H_p(2\omega)\right\rangle \xrightarrow{\text{PPKTP}} \left|H_s(\omega)\right\rangle \left|V_i(\omega)\right\rangle. \tag{9}$$

Here, the subscripts $p$ , $s$ , and $i$ denote the pump, signal, and idler photons, respectively, and $\omega$ denotes the central frequency. Based on the SPDC process, polarization-entangled photon pairs can be generated by constructing an interference superposition of two orthogonal polarization components of the pump laser. As shown in Fig. 2a, the pump laser is tailored to the desired polarization state by a combination of a PBS, a HWP, and a QWP, and is subsequently reflected by a dichroic mirror (DM) into a Sagnac interferometer. Here, the Sagnac interferometer is composed of a pair of off-axis parabolic mirrors (OAPMs), a dual-wavelength half-wave plate (DHWP), and a dual-wavelength polarization beam splitter (DPBS). The symmetric OAPMs serve to focus and collimate the photons, thereby enhancing the utilization efficiency of the pump photons and the collection efficiency of the down-converted photons. Inside the Sagnac interferometer, the horizontal and vertical linear polarization components of the pump laser propagate in the clockwise (CW) and counter-clockwise (CCW) directions, respectively. Because the optical axis of the DHWP is set at $45°$, the quantum state transformation processes experienced by the two orthogonally polarized pump components can be respectively expressed as:

$$\begin{aligned} &\text{CW:} \left|H_p(2\omega)\right\rangle \xrightarrow{\text{PPKTP}} \left|H_s(\omega)\right\rangle \left|V_i(\omega)\right\rangle \xrightarrow{\text{DHWP}} \left|V_s(\omega)\right\rangle \left|H_i(\omega)\right\rangle. \\ &\text{CCW:} \left|V_p(2\omega)\right\rangle \xrightarrow{\text{DHWP}} \left|H_p(2\omega)\right\rangle \xrightarrow{\text{DHWP}} \left|H_s(\omega)\right\rangle \left|V_i(\omega)\right\rangle. \end{aligned} \tag{10}$$

Therefore, when the polarization state of the pump laser is set to $\left|\psi\right\rangle = \cos\kappa\left|H_p\right\rangle + \sin\kappa e^{i\tau}\left|V_p\right\rangle$, the two-photon polarization state at the output of the Sagnac interferometer can be expressed as an interference superposition state arising from the two polarization components:

$$\left|\Psi\right\rangle = A\cos\kappa\left|H_s\right\rangle\left|V_i\right\rangle + B\sin\kappa e^{i(C+\tau)}\left|V_s\right\rangle\left|H_i\right\rangle. \tag{11}$$

Here, the parameters $A$ and $B$ denote the transmission efficiencies of the CW and CCW components of the Sagnac interferometer, respectively, and $C$ represents the phase difference accumulated between the CW and CCW components during propagation. Consequently, by tailoring the polarization state of the pump laser, the conditions $A\cos\kappa = B\sin\kappa$ and $C+\tau=0$ can be satisfied. Upon normalization, the signal and idler photons are prepared in the maximally entangled third Bell state:

$$\left|\Psi^+\right\rangle = \frac{1}{\sqrt{2}}\left(\left|H_s\right\rangle\left|V_i\right\rangle + \left|V_s\right\rangle\left|H_i\right\rangle\right). \tag{12}$$

**Fabrication of the metasurface and *q*-plate**

Both metasurface and $q$-plate are fabricated on quartz glass substrates using laser direct writing technology. The variation in laser intensity creates grating-like nanostructures within the initially isotropic glass, thereby producing a periodic modulation of the refractive index. This spatially inhomogeneous refractive-index distribution gives rise to the birefringence effect in the metasurface. The birefringent phase retardation can be expressed as $\Delta\varphi = 2\pi(n_e - n_o)d/\lambda$, where $d$ is the writing depth and $\lambda$ is the operating wavelength. In the linear approximation, the effective ordinary and extraordinary refractive indices are expressed as:

$$\begin{aligned} n_0 &= \sqrt{Fn_1^2 + \left(1-F\right)n_2^2}, \\ n_e &= \sqrt{\frac{n_1^2 n_2^2}{Fn_2^2 + \left(1-F\right)n_1^2}}. \end{aligned} \tag{13}$$

Here, $F$ is the fill factor, and $n_1$ and $n_2$ are the refractive indices of the two media constituting the nanogratings. Consequently, the metasurface and $q$-plate exhibit spatially uniform birefringent phase retardation with an anisotropic optical axis.

**External trigger mode of the ICCD**

In the external trigger mode of the ICCD, the opening and closing of its shutter are controlled by an external electrical signal. Since the signal and idler photons are always generated in pairs simultaneously in the SPDC process, they can serve as trigger signals for each other. Taking the idler photon as the trigger signal as an example: it is first converted into an electrical signal by an SPCM and then transmitted to the trigger port of the ICCD, causing the shutter to open for a duration of $\tau = 4$ ns. By means of fiber-optic delay lines and precise configuration of the ICCD software, it can be ensured that the signal photon reaches the photosensitive sensor of the ICCD exactly within this time window $\tau$. Consequently, accurate time-coincidence imaging can be realized through the external trigger mode of the ICCD.

**Acknowledgements.** This work was supported by the National Natural Science Foundation of China (Grants No. 12174097). Singapore Ministry of Education (MOE) AcRF Tier 1 (Grants Nos. RG157/23 and RT11/23), Singapore Agency for Science, Technology and Research (A*STAR) (Grants Nos. M24N7c0080 and H25-MRO3489), and Nanyang Assistant Professorship Start Up grant.

**Competing interests.** The authors declare no competing interests.

**Contributions.** Y.W. conducted the experimental measurements, prepared the figures and tables, and wrote the first draft of the manuscript. Z.G. optimized the experimental parameters and performed the data analysis. M.L. participated in the experimental setup. Y.S. supervised this project. H.L. conceived the initial idea and supervised this project. All authors contributed to finalizing the manuscript.

**Corresponding author.** Correspondence to Yijie Shen or Hailu Luo.

**Data availability.** Data underlying the results presented in this paper may be obtained from the authors upon reasonable request.